\documentclass[journal=jacsat,manuscript=article]{achemso}

\usepackage[version=3]{mhchem} 
\setkeys{acs}{maxauthors=90}
\usepackage{amsmath, amsfonts, mathtools, amssymb}
\usepackage{here}
\usepackage{hyperref}
\usepackage{xcolor}
\usepackage{caption}
\usepackage{subcaption}
\usepackage{microtype}
\usepackage{soul}

\author{Taoufik Sakhraoui}
\email{taoufik.sakhraoui@osu.cz}
\affiliation{Department of Physics, Faculty of Science, University of Ostrava, 701 03 Ostrava, Czech Republic}
\author{František Karlický}
\affiliation{Department of Physics, Faculty of Science, University of Ostrava, 701 03 Ostrava, Czech Republic}

\title[]{DFTB coupled with NEGF study of the structural, electronic and transport properties of goldene 2D material}

\abbreviations{IR,NMR,UV}

\begin{document}

\begin{abstract}
We report the structural, electronic, and transport properties of the goldene 2D material using the density functional tight-binding (DFTB) method. Electronic transport calculations were conducted in conjunction with the non-equilibrium Green’s functions (NEGF) technique. Our study reveals that the Au 2D material is dynamically and thermally stable, and it possesses good elastic properties. On the other hand, goldene has a linear relationship between current and voltage at low potentials, indicating its metallic character. The calculated current-potential curve correlates  well with transmission functions and the electronic density of states around the Fermi level. \\
We also investigated the electronic structure and magnetic properties of silicon (Si)-doped Au 2D material. Our results show that the Si atom can induce a local magnetic state in the goldene monolayer. The resulting magnetic moment is 0.63 $\mu_{\rm B}$.
\end{abstract}

\section{Introduction}

Over the past two decades, significant interest has been devoted to two-dimensional (2D) materials due to their unique physical and chemical properties. 2D compounds are used in electronic devices, sensors, and energy storage \cite{Jiaqi2024, Liu2020, Hassan2023, Dai2024}. Before the successful synthesis of goldene in 2024 \cite{Kashiwaya2024, Kashiwaya2025}, significant attempts were made to create 2D gold materials. A first-time preparation of free-standing, one-atom-thick 2D goldene was reported in 2022 \cite{Sharma2022}. Authors show, using High-resolution transmission electron microscopy (HRTEM) that goldene possesses a honeycomb lattice and a semiconductor response with a knee voltage of around 3.2 V.S. Forti {\it et al.} synthesized a large-area semiconductor 2D goldene single layer, stabilized between SiC and graphene \cite{Forti2020}. Interestingly, the manipulation of the amount of gold at the SiC/graphene interface leads to a semiconductor to metal transition in the 2D graphene. W. Yuan \cite{Yuan2020} prepared a single layer of goldene sandwiched between graphene layers. The authors in the following reference \cite{YiCui2024} epitaxially grew 2D gold nanodisks layers between MoS$_{{\rm2}}$ substrate and MoS$_{{\rm2}}$ film. Upon heating, the nanoparticles flattened into nanodisks.

After being synthesized as free-standing, several-atoms-thick layers, or monolayers confined on or inside templates \cite{WangL2015, WangX2019, Zhao2020, Bhandari2019}, the exfoliation of single-atom-thick gold (goldene) has been achieved through wet-chemical etching away Ti$_{{\rm3}}$C$_{{\rm2}}$ from nanolaminated Ti$_{{\rm3}}$AuC$_{{\rm2}}$, initially formed by substituting Si in Ti$_{{\rm3}}$SiC$_{{\rm2}}$ with Au at Linköping University in Sweden \cite{Kashiwaya2024, Kashiwaya2025}. The experimentally created 2D goldene has a hexagonal structure. M. L. Pereira, Jr {\it et al.} try to answer the question about goldene stacking structure \cite{Pereira2025}. They show that the  AA-like stacking of multilayers is maintained up to six layers; then, a bulk-like ABC-like stacking structure appears at seven layers, and finally, the multilayers converge to a bulk structure for more than ten layers. \\
After the 2D goldene was exfoliated, some groups attempted to use first-principles calculations to study the 1D nanotubes (GNT) and explore their structural and electronic properties \cite{Shota2024}.  Based on DFT, D. Li {\it et al.} \cite{DongboLi2025} reported a theoretical study on the structure, electrical, and optical properties of goldene, silverene, and copperene 2D materials. Their results reveal that goldene possesses a dynamically  stable 2D planar structure and exhibits high conductivity. S. Zhao {\it et al.} conducted DFT calculations on the conductivity of goldene from electron-phonon scattering; they found that 2D goldene shows a very high intrinsic conductivity at room temperature, which is in the same order of magnitude as that of lightly doped graphene and much larger than that of other 2D materials \cite{Zhao2024}. Therefore, goldene 2D material may be an excellent alternative to the 3D bulk as a conductor in electronic devices. The novel 2D material was theoretically used as an electrode to design metal/semiconductor heterojunctions with 2D (Mo, W)(S, Se) semiconductors \cite{Nguyen2025}. Goldene monosheet possesses metallic behavior, and its electronic properties may be tuned by applying strain, which makes it promising for integration in the metal/semiconductor heterojunctions as an electrode material \cite{Nguyen2025}. Additionally, first principles, molecular dynamics, and machine learning potential studies showed that goldene 2D material is structurally stable up to 700 K \cite{Mortazavi2024} and maintains its metallic nature. The lattice thermal conductivity of goldene was found to be around 10$\pm$2 W/(m K) \cite{Mortazavi2024}, which is five times higher than that of the bulk gold material (2 W/(m K) \cite{Jain2016} or 2.6 W/(m K) \cite{WangYan2016}). First principles calculations reveal that 2D goldene, unlike its three-dimensional bulk counterpart, shows a very high intrinsic conductivity at room temperature \cite{Zhao2024}. It preserves its conductivity under tensile strain and through structural vacancy defect creation \cite{Berdiyorov2026}, which makes it a conductor suitable for use in flexible electronics based on 2D materials.    

Recently, DFTB method was used to investigate the properties of gold-based materials. The authors of the following ref. \cite{Maghrebi_2023} show that it is possible to rapidly access the ground-state as well as the excited-state properties of Gold nanoclusters keeping a qualitative accuracy compared to DFT method. J. Morehouse et al. \cite{Morehouse_2026} evaluate the performance of DFTB+Grimme’s D3 (BJ) dispersion correction compared to DFT for gold nanocluster-amino acid complexes. They found that the used technique offers an adequately accurate framework for screening Au-biomolecule interactions. Regarding the conductivity of gold/goldene, we are not aware of any DFTB study of goldene based junction. However, this method was employed to represent a study of tunneling across gold/molecules/gold junction \cite{Khadijeh2017}. The authors show good agreement with experiments and DFT calculations. Further, DFTB method was also successfully used to investigate the transport properties of 2D MXene based junctions \cite{Khanal_2023}. The authors  validate the geometries and the electronic structure by comparing to DFT results calculated by VASP package. Moreover, we recently used the mentioned method to investigate the effect of defect and adsorption on the electronic and magnetic properties of 2D MXene materials \cite{taoufik2022, Taoufik2024}.

Although there has been significant research on 2D metals, particularly goldene, a thorough examination of the electronic transport, influence of external doping, and mechanical characteristics of goldene has not yet been extensively performed. Recently, goldene was shown to be a promising electrode in metal/semiconductor junction and it was integrated in the 2D Goldene/2D transition metal dichalcogenides heterojunctions \cite{Nguyen2025} by using first-principles calculations. Herein, for the first time, we exploit the advantages of DFTB method, i.e. relative low requirements for computing resources and high accuracy to investigate the structural, mechanical, electronic, and transport properties of the goldene 2D material. We also study the effect of Si doping in goldene. In addition, our molecular dynamics simulations show that goldene remains structurally stable up to 700 K. The dynamical stability was also proved by phonon dispersion. Moreover, we emphasized the goldene 2D material’s exceptional electronic conductivity.

\section{Theoretical details}
The DFTB method provides a good balance between accuracy and computational efficiency. It consists of a computational model that applies a tight-binding scheme to the DFT. In  spite of its good computational efficiency, the accuracy of the DFTB method depends on the so-called Slater-Koster parameter files, which should be appropriately benchmarked. In many cases, the DFTB results may reach the DFT-level accuracy \cite{Jianhua_2025, Selli_2017, Bai_2023}. Y. A. Cetin et al. \cite{Cetin_2022} compared two computational schemes (DFT and DFTB) and show that both converge to the same qualitative results and trends. Interestingly, P. Quaino et al. \cite{Quaino_2023} show that the DFTB method is better compared to DFT in treating the potential at metal-solution Interfaces. Moreover, Qi Wang et al. \cite{Wang_2017} show that the DFTB method is able to accurately predict the thermal transport properties in various Si systems, showing excellent computation efficiency and transferability. The DFTB is also used for the accurate calculation of the electrical conductivity of 2D material \cite{Dmitry_2025}. \\
All calculations were carried out using the DFTB method \cite{Elstner_1998, Frauenheim_2002, Cui2014} as implemented in the DFTB+ program package \cite{Aradi2007, Aradi2020} and the PTBP parameter set for the basic electronic Slater-Koster parameters \cite{Cui2024, Cui2024_1}. The DFTB approach enables the treatment of thousands of atoms at the quantum level, providing a thorough and careful prediction of intricate material structures while maintaining computational efficiency and satisfactory accuracy. For Brillouin zone integration, the k-point meshes were set to 12$\times$12$\times$1 for the Au 2D material, and only $\Gamma$ point was used for the Si-doped 7$\times$7$\times$1-Au supercell model. During the estimation of the self-consistent charge (SCC) parameters that control the electronic minimization, the self-consistent procedure stopped when the difference in any charge between two SCC cycles was lower than 1$\times$10$^{-6}$ eV/atom. NVT simulations were conducted at the same temperatures of 300 K, 500 K, and 700 K, using the Nose-Hoover thermostat \cite{Glenn1996}. \\
The DFTB+ package is interfaced with the phonopy code \cite{Togo2015}, which serves as a suitable tool  for the computation of the phonon dispersion using the supercell method. The calculation process is  organized as follows; first, phonopy creates several supercells with slightly displaced atoms.  Then,  DFTB+ computes the atomic forces through single-point calculations on these structures. Finally, Phonopy gathers this data to assess the force constants via numerical differentiation and then creates the dynamical matrix, which is subsequently used for diagonalization to determine the phonon band structure. Converged results were obtained by taking the supercell dimension to be (7$\times$7$\times$1) for the goldene 2D material. For all supercells, the DFTB single-point calculations were carried out at the $\Gamma$-point.\\
We compute the electronic transport properties of a goldene monolayer using the DFTB method with the non-equilibrium Green’s function (NEGF) \cite{Brandbyge2002, Datta1995} technique and the Landauer-Büttiker formalism \cite{LandauerButtiker1985}. The NEGF formalism, which is a precise and effective way to calculate transport properties, was implemented in the DFTB+ code by A. Pecchia et al. \cite{Pecchia_2008}. Numerous authors have successfully used the NEGF+DFTB approach \cite{Li2022, Khadijeh2017, Fatih2022}. The electrical current through the device under non-equilibrium conditions, i.e., for a finite bias voltage (V), is calculated using the Landauer formula \cite{Landauer1957}: $I(V)=\dfrac{2e}{h}\int_{\mu_L}^{\mu_R} [f(E-\mu_L) - f(E-\mu_R)]T(E, V)dE$, \\
where V is the applied bias voltage, $T(E, V)$ is the energy- and voltage-resolved transmission function, h is the Planck constant,  $f(E-\mu_{L/R})=\dfrac{1}{1+e^{(E-\mu_{L/R})/K_B\tau}}$ is the Fermi-Dirac distribution function at an electronic temperature $\tau$ of 300 K, $\mu_{L/R}$ are the chemical potentials of the left/right electrodes, respectively. $\mu_L$ and $\mu_R$ are defined as $\mu_L=\varepsilon_f$ + eV/2 and $\mu_R=\varepsilon_f$ - eV/2, wherein $\varepsilon_f$ is the Fermi energy of the electrodes and e is the elementary charge. The transmissions were obtained from the NEGF, following the implementation in the DFTB+ software \cite{Pecchia_2008}. \\
The simulated supercell consists of three parts: the left and right semi-periodic electrodes, separated by a central scattering region. The structures are optimized applying 3D periodic boundary conditions and total energies are converged to within 10$^{-5}$ eV. The default GammaFunctional solver was used to solve the Poisson equation. The calculations involve the self-consistent charge correction \cite{Elstner_1998}, which  takes into account the electron density redistribution due to interatomic interactions. For transport calculations, we use a dense k-point sampling of 50$\times$12$\times$1 for the electrodes and 1$\times$12$\times$1 for the central region, using the Monkhorst-Pack scheme \cite{Monkhorst1976}. We set the electrode temperature to 300 K. The entire device is periodic normal to the direction of transport. It should be noted that there is good agreement between DFTB and DFT results obtained for the transport results \cite{Ghorbani-Asl2015}. The DFTB transport code \cite{Aldo2002} relies on a tight-binding (TB) parametrization and, as such, has the advantage of being able to treat larger junctions. Further information about the DFTB+ code, and transport calculations in particular, can be found in the Ref. \cite{Aradi2020}.

\section{Results and discussion}
\subsection{Dynamical and thermal stability}
\begin{figure}[H]
\centering
\includegraphics[width=0.5\textwidth]{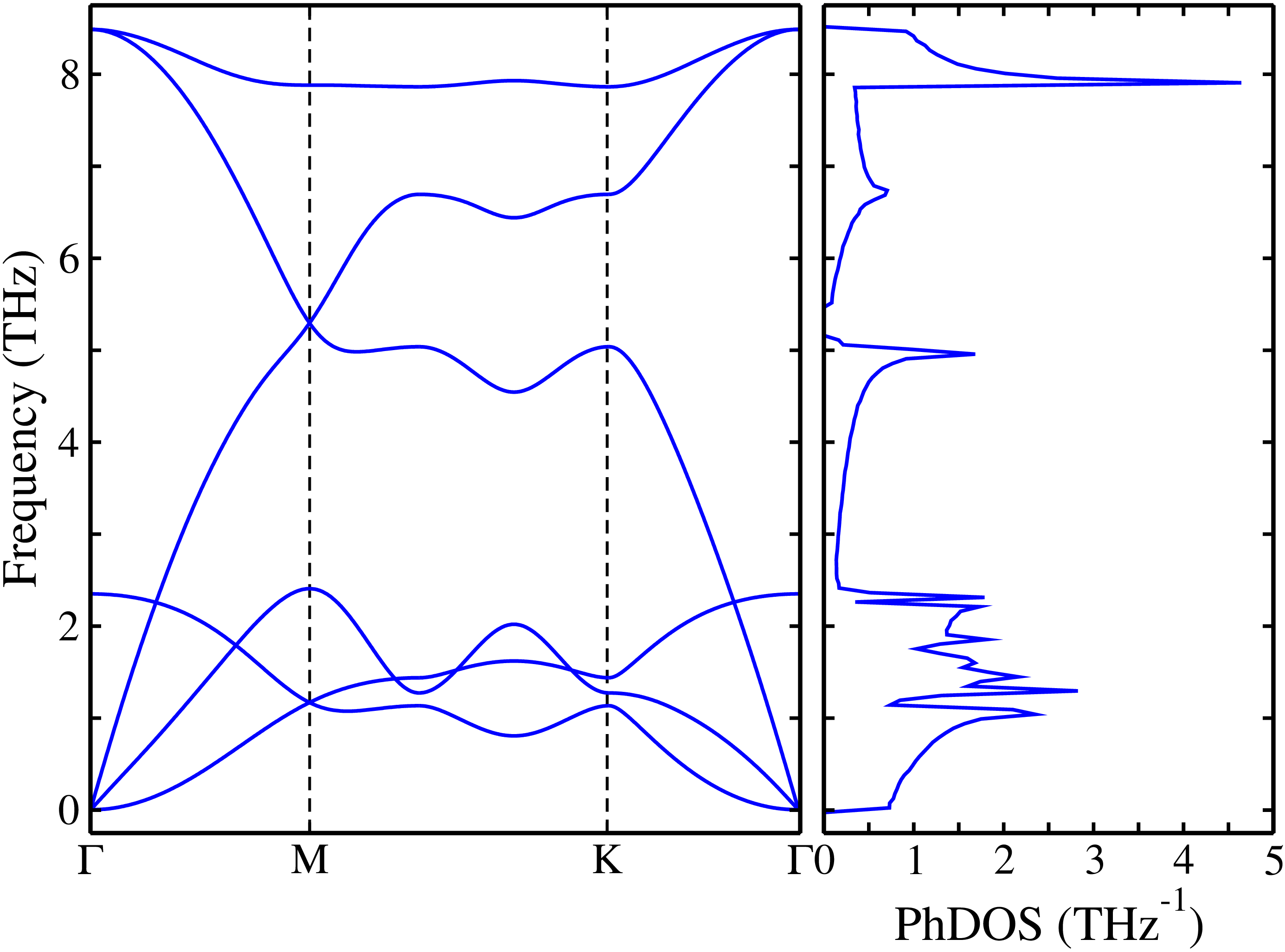}
\caption{Phonon dispersion curve and phonon density of states (PhDOS) of the goldene monolayer.}
\label{phon}
\end{figure}

\begin{figure}[H]
\centering
\includegraphics[width=0.8\textwidth]{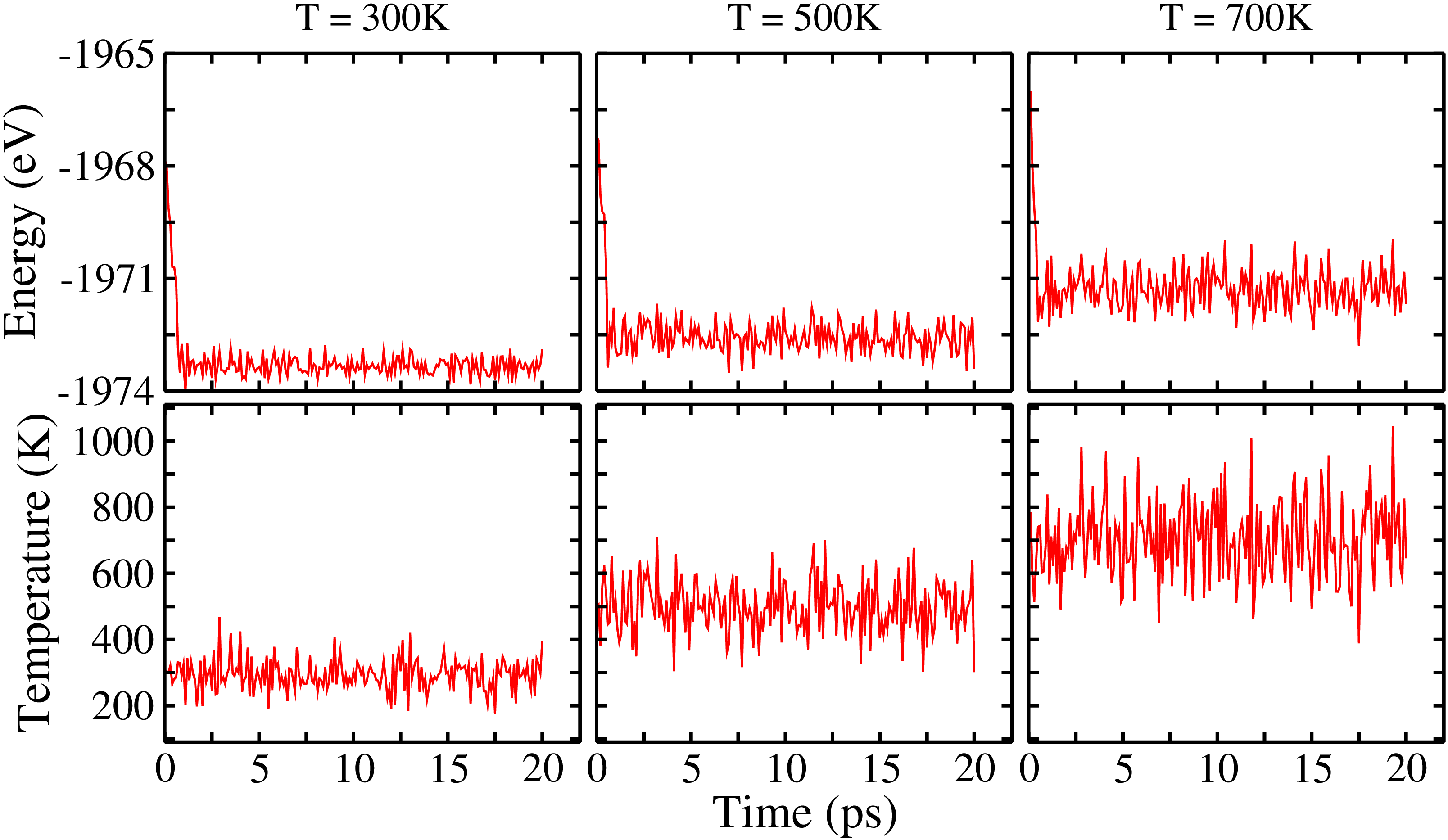}
\caption{The energy and temperature of the goldene monolayer as a function of simulation time at temperatures of 300 K, 500 K, and 700 K.}
\label{entempMD}
\end{figure}

One of the major concerns regarding the material under investigation is its dynamical stability.  In Fig. \ref{phon}, we plot the phonon spectrum to address this question. First of all, we clearly see that the Au 2D material exhibits positive vibrational energies for the entire  Brillouin zone, which confirms its dynamical stability. We also remark that the energy gap between the acoustic and optical modes is quite large.
In order to examine the thermal stability of the goldene material, we performed molecular dynamics (MD) calculations using 5$\times$5$\times$1 supercell. The NVT ensemble simulations were performed at 300 K, 500 K, and 700 K, for a total simulation time of 20 ps in time steps of 2 fs, using the Nose-Hoover thermostat \cite{Glenn1996}. The energy and temperature as a function of time steps are plotted in Fig. \ref{entempMD}. We notice that there is no deformation or bond breakage in the structure; therefore, the Au 2D material possesses good thermal stability. Moreover, the structure's overall energy remains almost constant. After the MD simulation is performed, the planar geometry of the goldene 2D material is retained at 300-700 K.\\
The remarkable thermal and dynamic stability of the goldene monolayer is verified by the recent study of B. Mortazavi \cite{Mortazavi2024}, where the author found that goldene maintains its dynamic stability under biaxial and uniaxial strains of 3\% and 6\%. 
To further evaluate the dynamical and thermal stability of goldene 2D material, we performed  phonon and MD calculations at 300 K for 20ps under strain from -4\% to +4\%. We also run MD calculation for the defective goldene material at 300 K, 500 K and 700 K. Results are shown in Figures S1, S2, S3 and S4. Notably, there is no observable imaginary frequency present in the phonon dispersion curves as shown in Fig. S3, suggesting that goldene is dynamically stable under strain. Moreover, the resulting energy/temperature fluctuations of the material over time, as shown in Fig. S1, S2, and S4 are very limited, suggesting that the goldene structure remains intact under strain and vacancy-defects. Therefore, goldene 2D material displays both dynamic and thermal stabilities under strain from -4\% to +4\%.
\subsection{Structural properties}
\begin{figure}[H]
\begin{subfigure}[b]{0.52\textwidth}
\includegraphics[width=\textwidth]{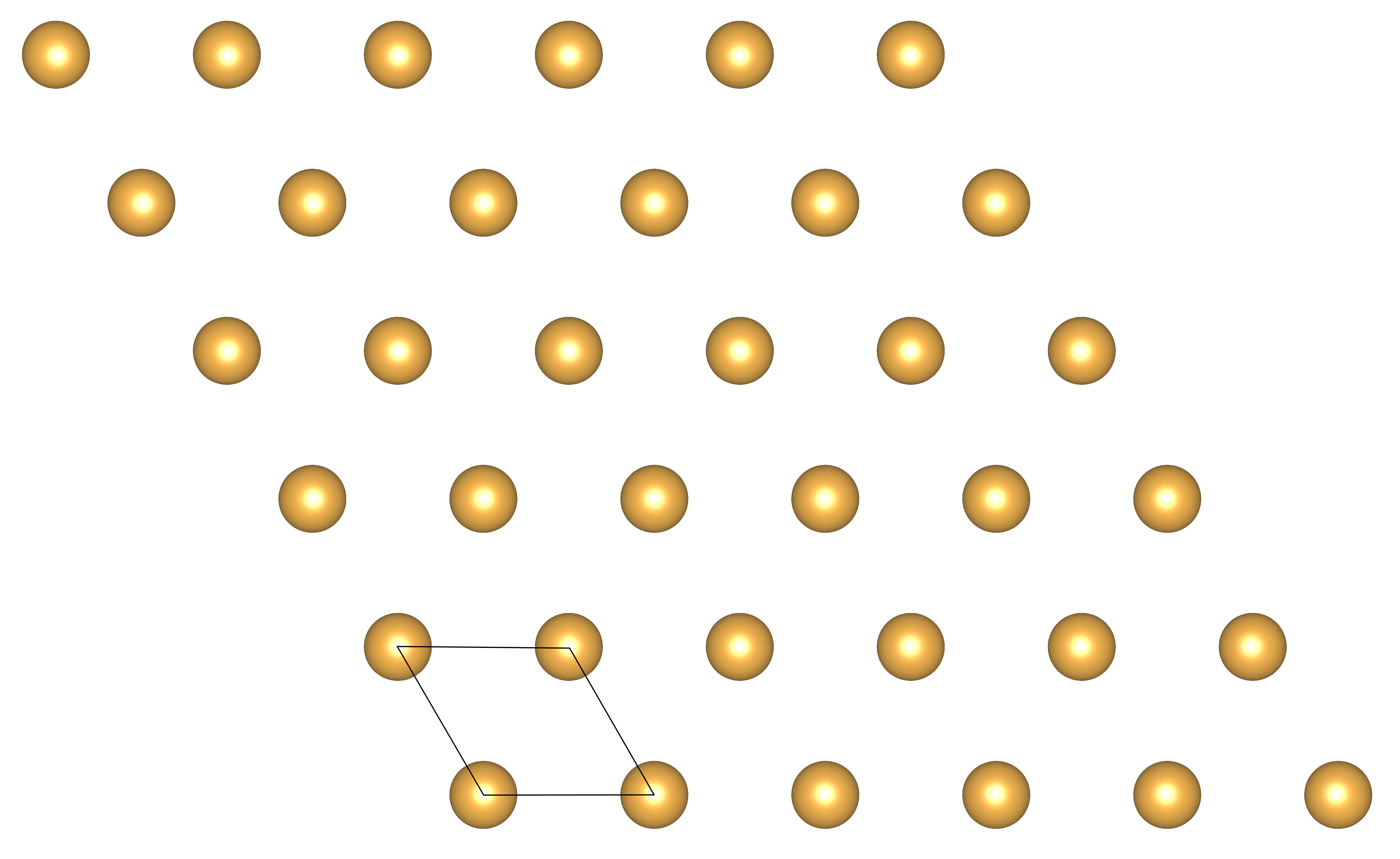}
\caption{ }\label{struct}
\end{subfigure}\hfill
\begin{subfigure}[b]{0.48\textwidth}
\includegraphics[width=0.7\textwidth]{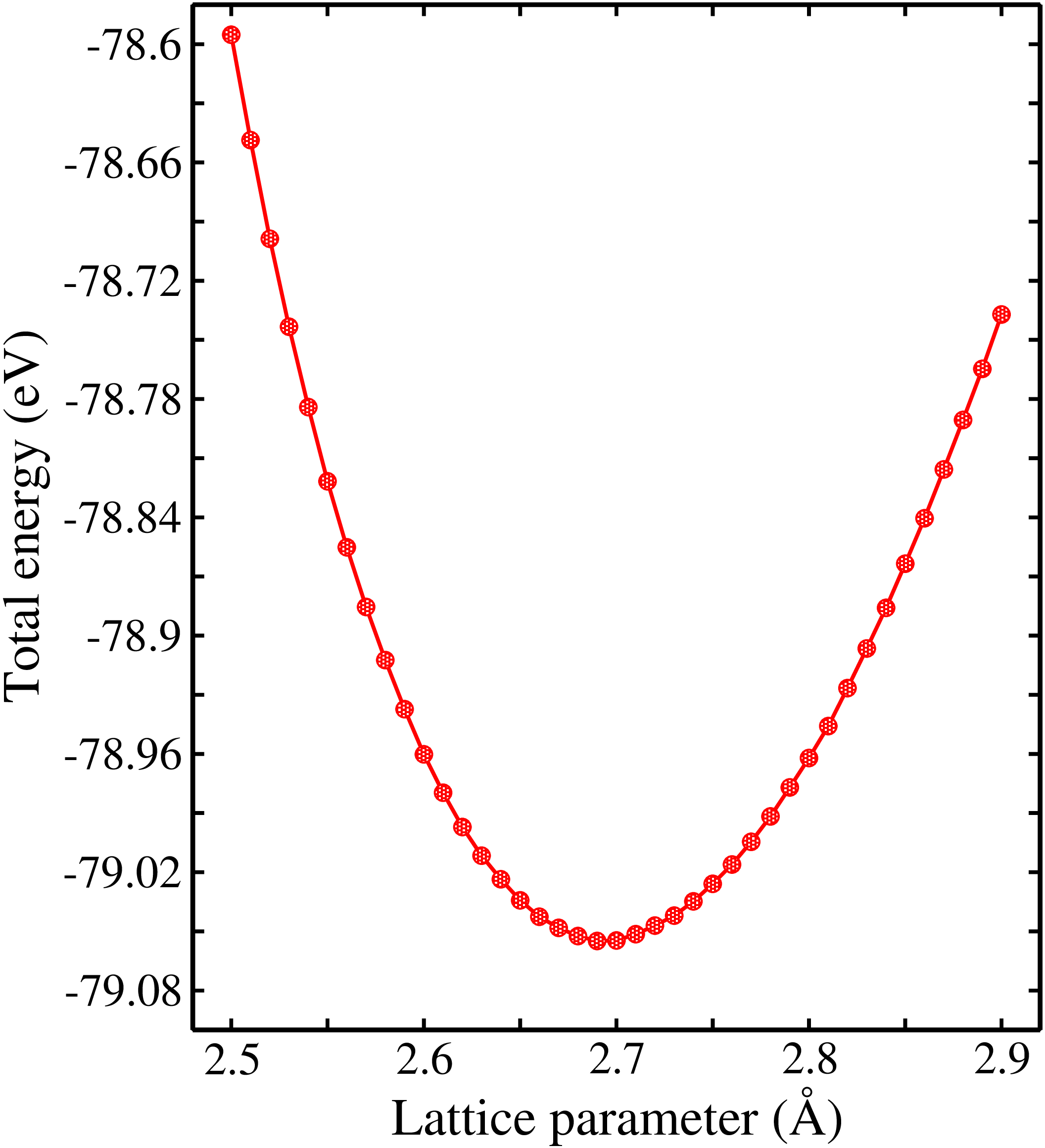}
\caption{}\label{latt}\end{subfigure}
\caption{ (a) Hexagonal close packed structure of Au monolayer. (b) Total energy as function as the lattice parameter. } \label{struct-latt}
\end{figure}

By conducting first-principles calculations, S. Ono shows that 2D goldene displays a planar hexagonal structure \cite{Ono2020}, and the lattice is  stabilized by metallic bonding. It corresponds to the top monolayer of the bulk Au(111) FCC lattice. The sheet of Au 2D material has P6/mmm (\#191) space group symmetry \cite{Sheremetyeva_2025}, as shown in Fig. \ref{struct-latt}. The unit cell contains a single Au atom. As a preliminary step, we optimize the equilibrium lattice constant for the goldene monolayer. In Fig. \ref{struct-latt}, we plot the total energy variation as a function of the lattice parameters, with full relaxation of all the atoms. We found an equilibrium lattice constant of 2.69 \AA. The DFTB calculated Au-Au bond length is 2.69 \AA. In comparison, our calculated equilibrium lattice parameter is in good agreement with the calculated (2.73 \AA \cite{Kashiwaya2024} and 2.75 \AA \cite{Yang2015}) and experimental (2.62 \AA \cite{Kashiwaya2024}) results.
We also calculated the cohesive energy of the Au 2D material to assess the stability of its structure, E$_{{\rm coh.}}$ = (-E$_{{\rm tot.}}$ + nE$_{{\rm Au}})/n$, where E$_{{\rm Au}}$ and E$_{{\rm tot.}}$ are the total energies of the Au atom and the Au monolayer, respectively, with n being the number of Au atoms per unit cell. \\
For comparison, the calculated DFTB cohesive energy of goldene is compared to a series of other 2D materials, including graphene, germanene, and silicene. The values of the corresponding cohesive energies are listed in Table \ref{coh-2D}.
\begin{table}[H]
\centering
\begin{tabular}{ccccccc}
\hline \hline
goldene& graphene& tetra-germanene& hex-agermanene& tetra-silicene& hexa-silicene \\
\hline
5.30 & 7.90 \cite{Shin2014}, 7.73 \cite{Ivanovskaya2010}&5.13 \cite{Nguyen2026} & 4.76 \cite{Nguyen2026} & 3.89 \cite{Qiao2017} & 3.90\cite{Qiao2017} \\
\hline 
\end{tabular} 
\caption{Cohesive energies in eV/atom of goldene (present DFTB work), graphene, germanene, and silicene.}
\label{coh-2D}
\end{table}

\subsection{Mechanical properties}
It has already been shown that the DFTB calculated elastic constants and moduli for bulk ZnO gave reasonable agreement with DFT results \cite{Chol2024}, which motivates us to consider the current method for the investigation of the mechanical properties of Au 2D material. Based on the matrix notation of Hooke’s law, the stress/strain relation is given by \cite{Zhang2010, Wei2014}: $\sigma = \varepsilon{\bf C}$,\\
where {\bf C} is a $6\times6$ tensor of elastic constants with elements $c_{ij}$ expressed in Voigt notation: 11=xx, 22=yy, 33=zz, 23=yz, 13=xz, 12=xy \cite{Andrew2012}. The form of the tensor depends on the symmetry of the lattice type of the material \cite{Nye1985, Yunguo2022}. In the case of a hexagonal lattice, there are five independent $c_{ij}$ elements $c_{11}$, $c_{12}$, $c_{13}$, $c_{33}$, and $c_{44}$. However, since goldene is a 2D material, this number is reduced to two independent elastic constants, i.e., $c_{11}=c_{22}$ and $c_{12}=c_{21}$, with $c_{66} = (c_{11} - c_{12})/2$ \cite{Mazdziarz2019, LiXue2024}. The elastic stiffness constants for the goldene material can be written as: 
\begin{center}
${\bf C} = \begin{pmatrix}
c_{11} & c_{12} & 0 \\
c_{12} & c_{11} & 0 \\
0      & 0      & c_{66}
\end{pmatrix}$ 
\end{center}
The Poisson’s ratio ($\nu$), in-plane Young’s modulus (Y), and shear modulus (G) are calculated from the elastic constants as follows \cite{Andrew2012}: $\nu$=C$_{12}$/C$_{11}$, Y=(C$_{11}^{2}$-C$_{12}^{2}$)/C$_{11}$, and G=C$_{66}$, respectively. In terms of the resistance of a sheet to stretching, the 2D layer modulus for hexagonal 2D materials is calculated by the following relation \cite{Andrew2012}: $\gamma$ = (C$_{11}$+C$_{12}$)/2. The calculated values of C$_{11}$, C$_{12}$, and C$_{66}$, Y, $\gamma$, G and $\nu$ are presented in Table \ref{latt-mech}.  

\begin{table}[H]
\centering
\begin{tabular}{lccccccc}
\hline \hline
material                           & C$_{{\rm11}}$ & C$_{{\rm12}}$ & C$_{{\rm66}}$ & $\nu$ & Y      & G      & $\gamma$ \\
\hline
Goldene (this work)                & 351.31        & 116.35        & 117.48        & 0.33  & 312.77 & 117.48 & 233.83 \\
graphene \cite{Singh2018}          & 358.9         & 65.1          & 146.9         & 0.18  & 347.1  & 146.9  & 212.0  \\
graphene \cite{Changgu2008} (Exp.) &   -           &   -           &     -         & 0.17  & 342    &  -     & -      \\
MoS$_2$ \cite{Singh2018}           & 132.3         & 32.8          & 49.5          & 0.25  & 124.14 & 49.5   & 82.5   \\
hBN \cite{Solajic2024}             & 290.77        & 63.93         & 113.42        & 0.22  & 276.41 & 113.42 & 177.35 \\
SiC \cite{Andrew2012}              & 179.7         & 53.9          & 62.9          & 0.30  & 163.5  & 62.9   & 116.8  \\
WO$_2$ \cite{Cakir2014}            & 261.2         & 87.8          & 86.7          & 0.34  & 86.7   & 231.7  & 174.5  \\
\hline 
\end{tabular} 
\caption{The elastic properties for goldene 2D material (i.e., C$_{{\rm11}}$, C$_{{\rm12}}$, C$_{{\rm66}}$, shear modulus (G), layer modulus ($\gamma$), Young’s modulus (Y) in the unit of N/m, and Poisson’s ratio ($\nu$) is dimensionless) compared to some other 2D materials from literature.}
\label{latt-mech}
\end{table}
Our calculated results show that the goldene monolayer represents very good elastic properties. The results are compared to a series of 2D materials. Theoretical values of some 2D materials are provided to serve as a basis for comparison. To the best of our knowledge, there are no experimental results on the elastic properties of goldene 2D material. We evaluate the in-plane Young modulus (or in-plane stiffness) to assess its mechanical stability. The calculated Young modulus of the Au 2D material is found to be 312.77 N/m, which is higher than that of a list of 2D materials, including MoS$_2$ (124.14 N/m \cite{Singh2018}), WO$_2$ (86.7 N/m \cite{Cakir2014}), hBN (276.41 N/m\cite{Solajic2024}) and SiC (163.5 N/m \cite{Andrew2012}) and slightly lower than graphene (347.1 N/m \cite{Singh2018}, 342.0 N/m \cite{Changgu2008}). This result indicates that the Au 2D material possesses high resistance to unidirectional compression as well as stretching, which confirms the work of B. Mortazavi et al. \cite{Mortazavi2024}. Goldene 2D material is of interest owing to its outstanding mechanical properties. This is scientifically important, as it means that the material can undergo significant deformation without stretching or fracturing. Therefore, it may be used for the fabrication of electronic devices without collapsing. Moreover, the electronic and magnetic properties are sensitive to strain. Thus, the mechanical robustness of goldene allows for higher strain sustainability and elastic deformation (not destructive).
\subsection{Electronic structure}
\begin{figure}[H]
\includegraphics[width=0.5\textwidth]{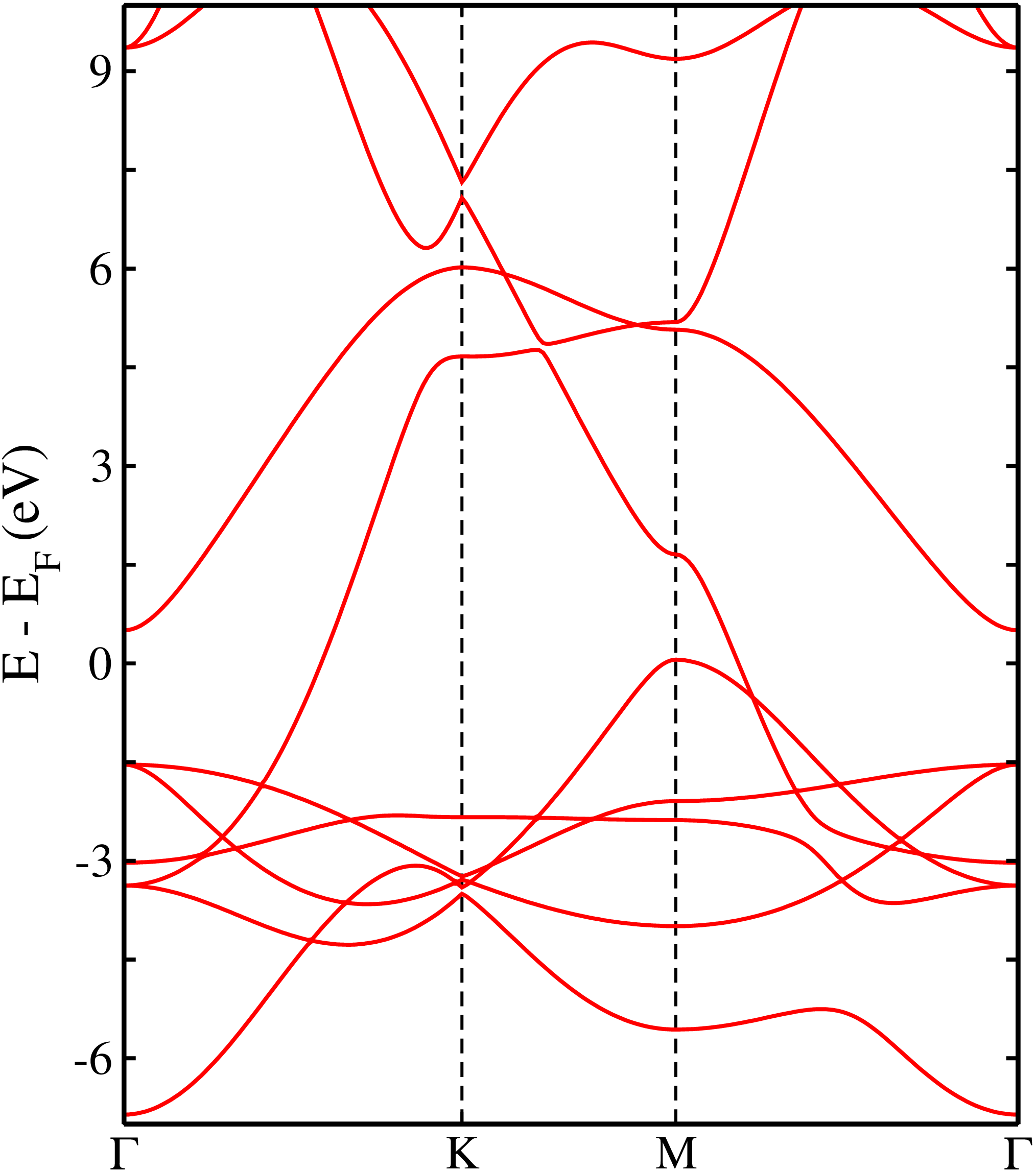}
\caption{Electronic band structure of goldene 2D material calculated by DFTB method.}
\label{band-Au_2D-dftb}
\end{figure}

We compute the electronic band structure of the Au 2D material, as shown in Fig. \ref{band-Au_2D-dftb}. We found that, similar to the FCC lattice of the gold 3D material \cite{Rangel2012}, the Au 2D material shows a band structure typical of metals. The Fermi level is located within the bands, and no band gap was observed at this energy. Comparing the DFT band structure \cite{Yang2015, Nguyen2025}, our DFTB calculations provide an accurate band structure. We clearly see a very similar shape of the band structure, especially near the Fermi level. G. R. Berdiyorov et al. show that the 2D goldene retains its metallic behavior under bending, twisting, mechanical strain, and vacancy defect creation \cite{Berdiyorov2026}. We note that we also tested the possibility of magnetism in the Au 2D material by performing spin-polarized calculations, and we found that, similar to the bulk \cite{Rangel2012}, the goldene monolayer is a nonmagnetic metal.

\subsection{Electronic transport properties}
\begin{figure}[H]
\centering
\includegraphics[width=0.99\textwidth]{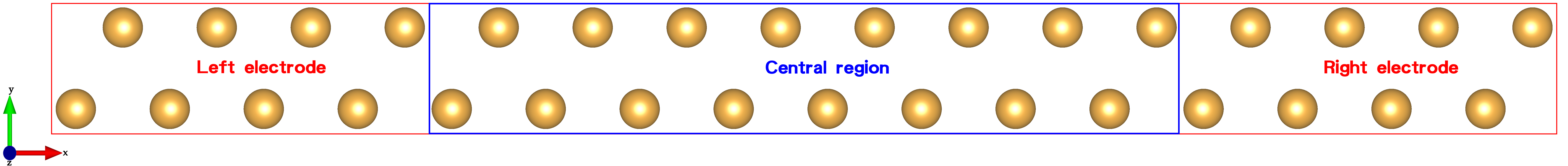}
\caption{Setup of the Au/Au/Au junction for the transport calculations, which consists of the right and left electrodes, and the central region.}
\label{transp-model}
\end{figure}
\begin{figure}[H]
\centering
\includegraphics[width=0.6\textwidth]{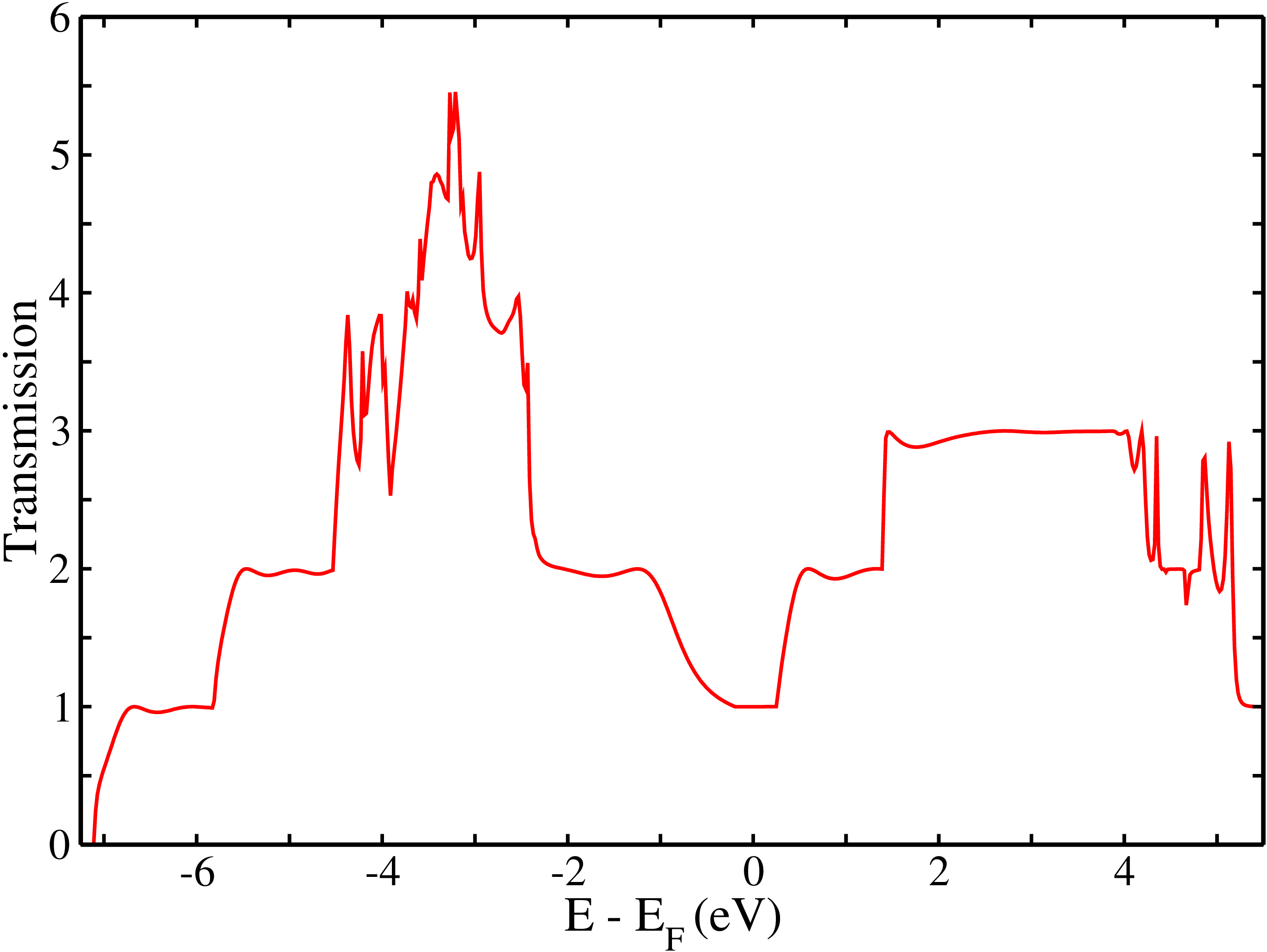}
\caption{The calculated (zero-bias) transmission function for the goldene 2D material. The Fermi level is shifted to zero.}
\label{tranmission-goldene}
\end{figure}

The considered system for the study of the transport properties consists of left and right electrodes separated by the central region (device). We use the optimized unit cell structure to construct the two-probe device configuration from pristine planar goldene. The structure is shown in Fig. \ref{transp-model}. We begin by investigating the transport properties of goldene at equilibrium (without bias voltage). We plot the transmission spectrum, T(E), at 0 V in Fig. \ref{tranmission-goldene}. 
The goldene is metallic in nature, which is confirmed by the absence of a transmission gap around the Fermi level. We see that there are multiple peaks in the energy range, which are associated with the existence of multiple transmission channels in that region. The transmission spectrum near the Fermi level (E$_{\rm F}$) showed a non-zero value, which indicates that there is direct tunneling from one electrode to the other, causing transmission near the Fermi level. The strong transmission spectra around E$_{\rm F}$ create good conduction from the Au/Au/Au junction. We note that the transmission coefficients around the Fermi level are higher than gold/molecule/gold \cite{Ke_2007, Strange_2008}, gold chain \cite{Calzolari_2004}, and gold/atom Au/gold \cite{Faleev_2005}.

\begin{figure}[H]
\begin{subfigure}[b]{0.49\textwidth}
\includegraphics[width=0.9\textwidth]{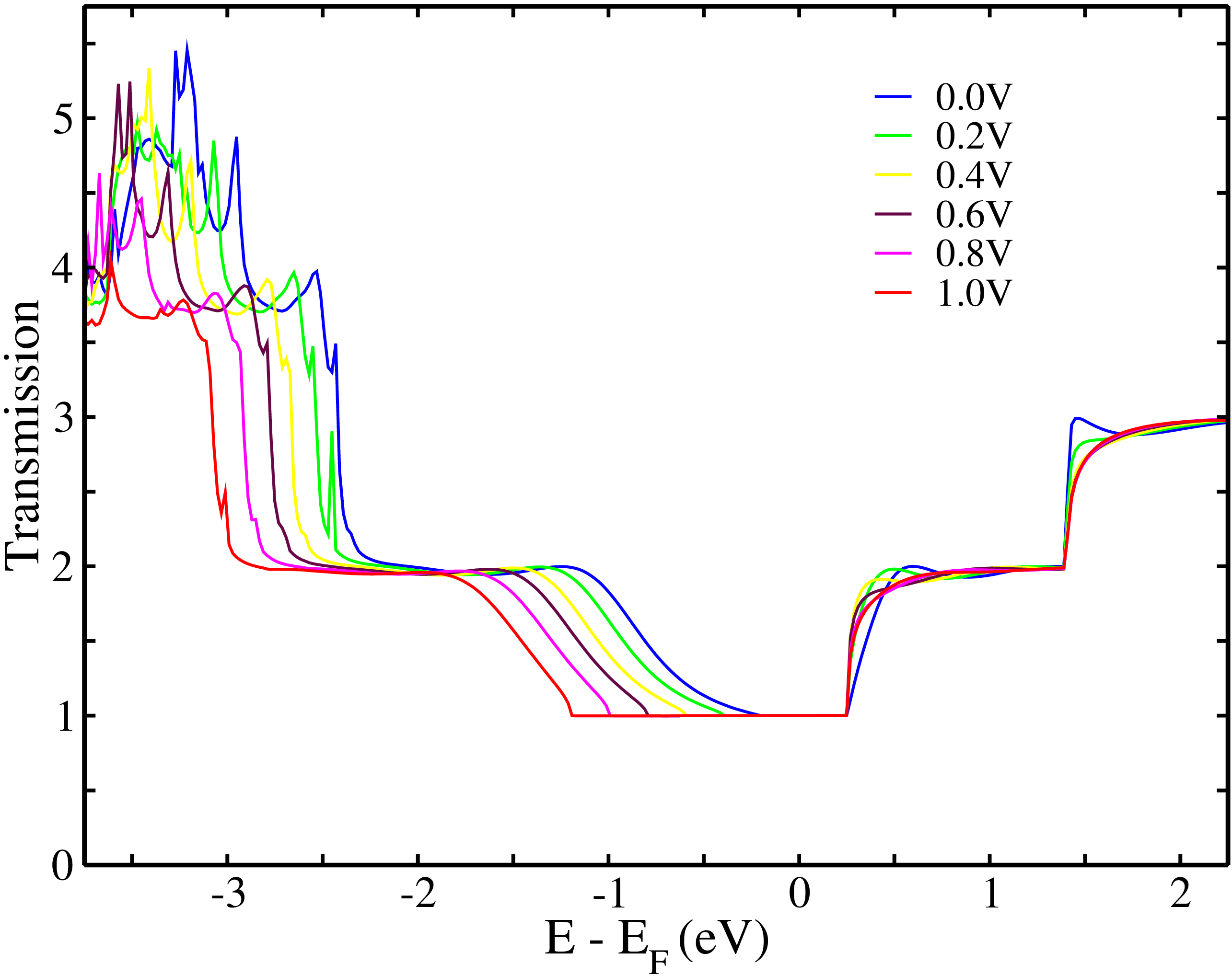}
\caption{}
\end{subfigure}\hfill
\begin{subfigure}[b]{0.49\textwidth}
\includegraphics[width=0.99\textwidth]{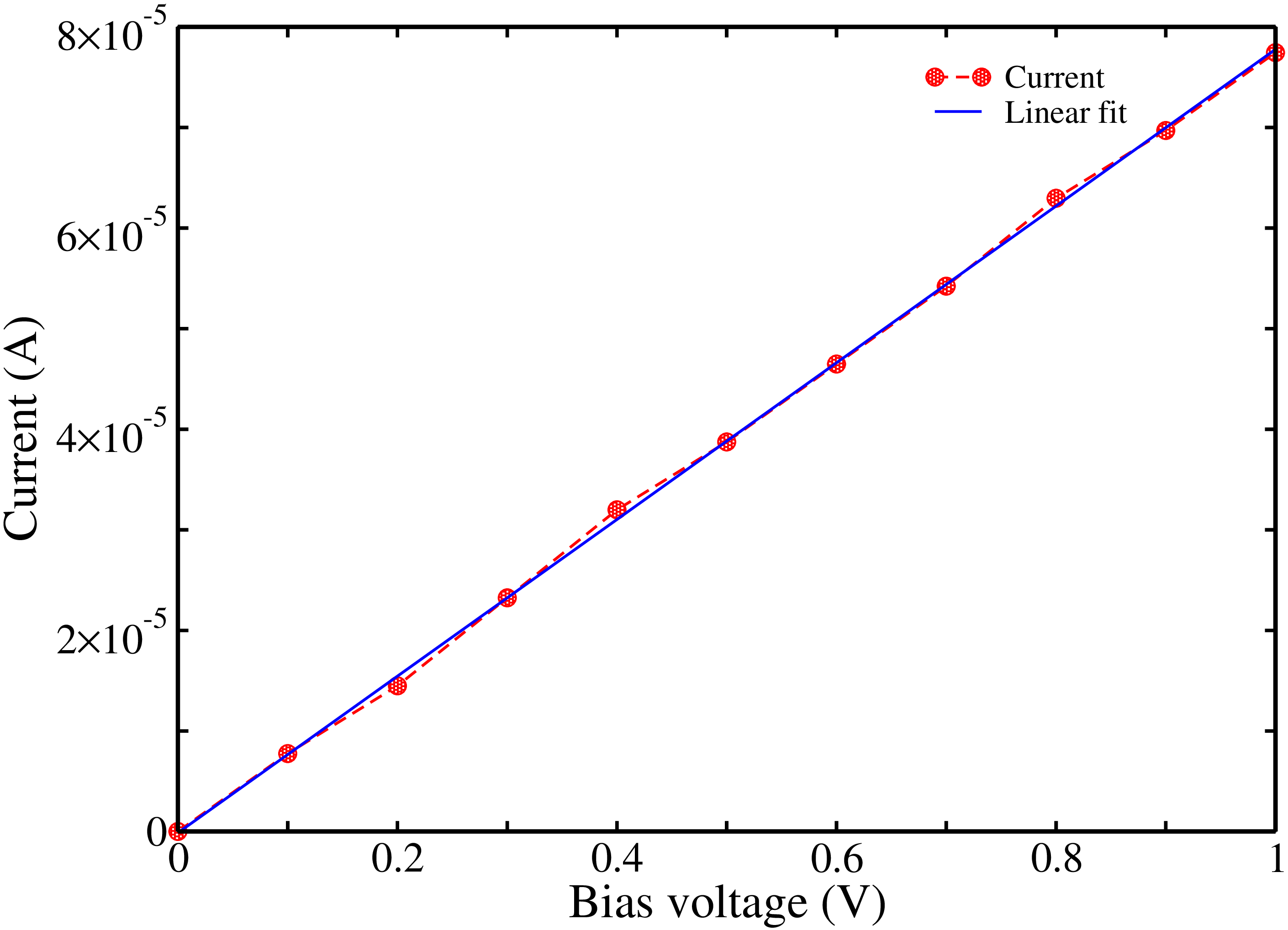}
\caption{}
\end{subfigure}
\caption{(a) The calculated transmission spectra under bias voltage. The Fermi level is shifted to zero, and (b) Current–voltage (I-V) characteristics of goldene monolayer.}
\label{currentvoltage}
\end{figure}

Next, we study bias voltage dependent current (I-V) characteristics, which present an important non-equilibrium transport characteristic. The I-V curve for the Au 2D material is plotted in Fig. \ref{currentvoltage}. We clearly observe that the I-V curves display typical metal behavior. The I-V curve shows Ohmic behavior, such that the current increases linearly with the increasing applied bias-voltage. This linear dependence was also found for the gold 3D material \cite{Faleev_2005}. First-principles calculations and Non-Equilibrium Green Function (NEGF) approach calculations on the electronic transport properties of 1D gold nanowires (AuNW) \cite{Bhandari2021} and goldene \cite{Berdiyorov2026} reveal that the I-V relations follow Ohmic behavior.

For further investigation of current-voltage characteristics, we examine the transmission spectra at different bias voltages ranging from 0.1 V to 1 V. Small increases in the nearly flat transmission near the Fermi level were observed with the increase in bias voltage, which is due to the suppression of electronic states close to the Fermi level while maintaining the conducting properties. However, there are still peaks not far from the Fermi level. These peaks in T(E) contribute directly to the current.\\
It is noteworthy that transmission characteristics and the I-V behavior are system dependent. High transmission probabilities at E$_{\rm F}$ are found for Co/MoS$_2$/Co \cite{Zhang_2016}, Cu-Fe$_3$GeTe$_2$-Cu system \cite{Lin_2020} and various MXene compositions, including Ti$_3$C$_2$, Ti$_3$C$_2$O$_2$, and Ti$_3$C$_2$(OH)$_2$ \cite{Khanal_2023}. The I-V behavior shows linear behavior for Ti$_3$C MXene \cite{Khanal_2023}, Cu-Fe$_3$GeTe$_2$-Cu system \cite{Lin_2020}, and Co/graphene/NiFe junctions \cite{Asshoff_2017}, nearly linear behavior for Cu-Fe$_3$GeTe$_2$-Cu system \cite{Lin_2020}, Ni(100)/BP/Ni(100) \cite{Chen_2016} and Fe/MoS$_2$/Fe junctions \cite{Dolui_2014}, and non-linear functions for CrI$_3$-based  systems \cite{Yan_2020}, Ni(111)/graphene \cite{Wu_2014}, and Ti$_2$CO$_2$ MXene based junctions \cite{Zhou_2016, Erdem_2019}.

\subsection{Si doping}

The pristine goldene monolayer is modeled by a 7$\times$7$\times$1 supercell with a lateral size of 18.83$\times$18.83 \AA$^2$, which contains 49 atoms. To dope the goldene 2D material, we follow the two-step process as suggested by the electron-beam-mediated substitutional doping scheme. First, we create a single Au-vacancy in the 7$\times$7$\times$1 goldene supercell and then incorporate the Si dopant into the vacancy site. The vacancy formation energy of 1 Au defect (E$_{{\rm form-vac}}$) is defined as E$_{{\rm form-vac}}$ = -E$_{{\rm goldene-free}}$ + E$_{{\rm vac}}$ + E$_{{\rm Au}}$, 
where E$_{{\rm goldene-free}}$, E$_{{\rm vac}}$, and E$_{{\rm Au}}$ are the total energies of goldene without a vacancy defect, defected Au 2D material, and the isolated Au atom. \\
The binding energy (E$_{{\rm b}}$) is calculated as: E$_{{\rm b}}$ = -E$_{{\rm vac}}$ + E$_{{\rm vac+Si}}$ - E$_{{\rm Si}}$. E$_{{\rm vac+Si}}$ and E$_{{\rm Si}}$ are the total energies of the defective goldene with a substitutional Si dopant and the energy of the isolated Si dopant, respectively. 
\begin{figure}[H]
\centering
\includegraphics[width=0.55\textwidth]{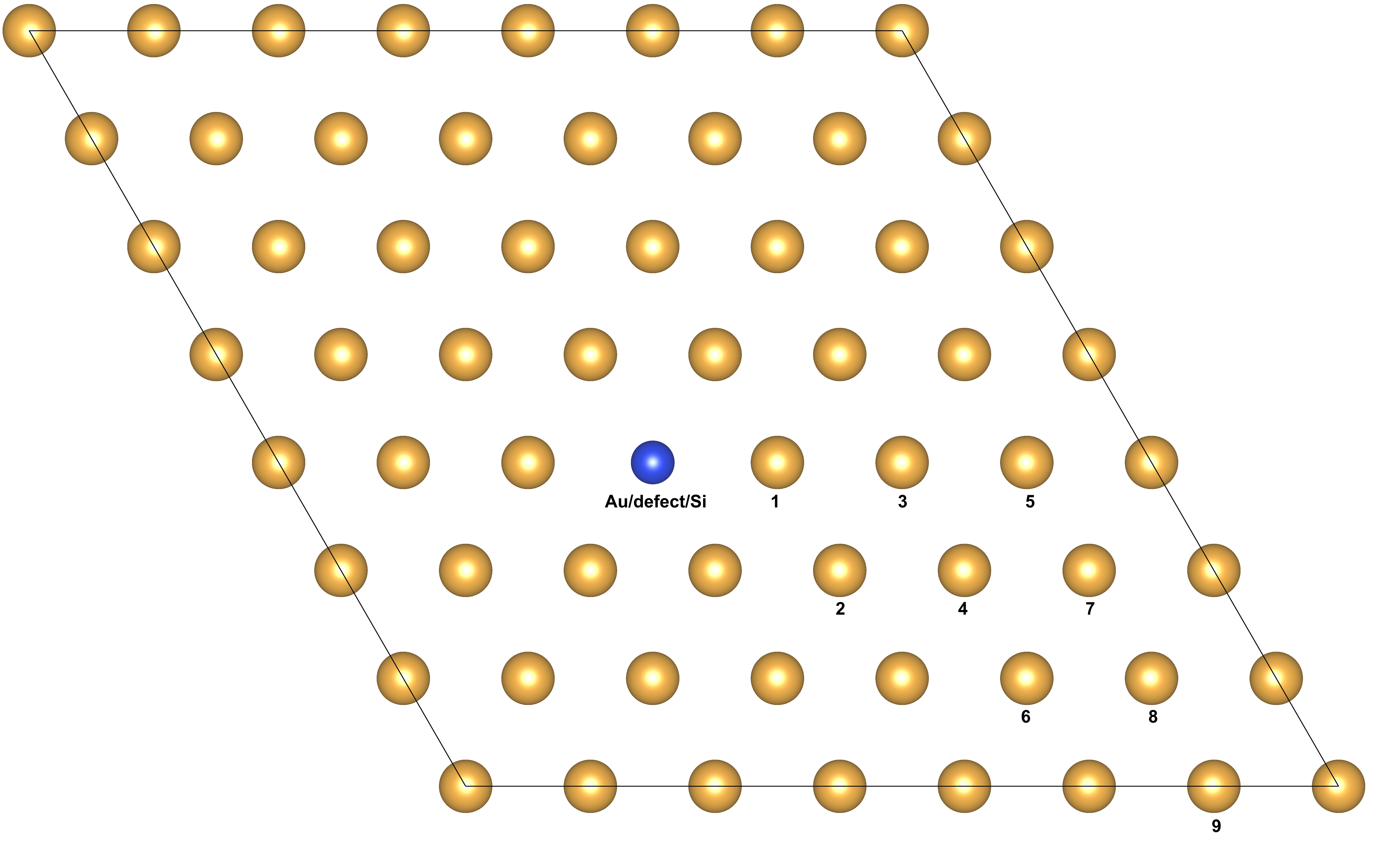}
\caption{Model of Si substitution on the surface of Au 2D material. The numbers represent the 1$^{{\rm st}}$, 2$^{{\rm nd}}$, 3$^{{\rm rd}}$, 4$^{{\rm th}}$, 5$^{{\rm th}}$, 6$^{{\rm th}}$, 7$^{{\rm th}}$, 8$^{{\rm th}}$, 9$^{{\rm th}}$ nearest neighbor (NN) to the considered Au atom or the vacancy defect or the Si atom.  }
\label{subs-NN}
\end{figure}
The Au vacancy defect is replaced by silicon. All atoms were relaxed, the bond lengths between the Si atom and the 1$^{{\rm st}}$, 2$^{{\rm nd}}$, 3$^{{\rm rd}}$, 4$^{{\rm th}}$, 5$^{{\rm th}}$, 6$^{{\rm th}}$, 7$^{{\rm th}}$, 8$^{{\rm th}}$, and 9$^{{\rm th}}$ nearest neighbors (NN) were calculated and compared to those in pure and defected structures. The results are listed in the table \ref{dist-subs}. 

\begin{table}[H]
\centering
\begin{tabular}{lccccccccc}
\hline \hline
               & 1    & 2    & 3    & 4    & 5    & 6    & 7    & 8     & 9     \\
\hline
Pure           & 2.69 & 4.66 & 5.38 & 7.11 & 8.07 & 9.31 & 9.70 & 11.72 & 13.98 \\
Au vacancy     & 2.78 & 4.61 & 5.62 & 7.26 & 8.22 & 9.57 & 9.91 & 11.98 & 14.30 \\
Si sustitution & 2.83 & 4.68 & 5.60 & 7.25 & 8.26 & 9.57 & 9.82 & 11.97 & 14.30 \\
\hline 
\end{tabular} 
\caption{Variation of the distance between the Au atom (pure structure), vacancy defect (1 Au defect) and Si atom (Si-substitution) and the 1$^{{\rm st}}$, 2$^{{\rm nd}}$, 3$^{{\rm rd}}$, 4$^{{\rm th}}$, 5$^{{\rm th}}$, 6$^{{\rm th}}$, 7$^{{\rm th}}$, 8$^{{\rm th}}$, 9$^{{\rm th}}$ nearest neighbors. Please refer to the Fig. \ref{subs-NN}}
\label{dist-subs}
\end{table}

In a doped sheet, Si connects with the three nearest Au atoms through sp$^{\rm2}$ hybridization. Si-doping results in a notable modification of the bond lengths with the surrounding atoms. All bond lengths of Si and Au NN atoms are higher than those in the pure system. Thus, little compression in Au-Au bonds occurs in the doped system. 
\begin{figure}[H]
\centering
\includegraphics[width=0.7\textwidth]{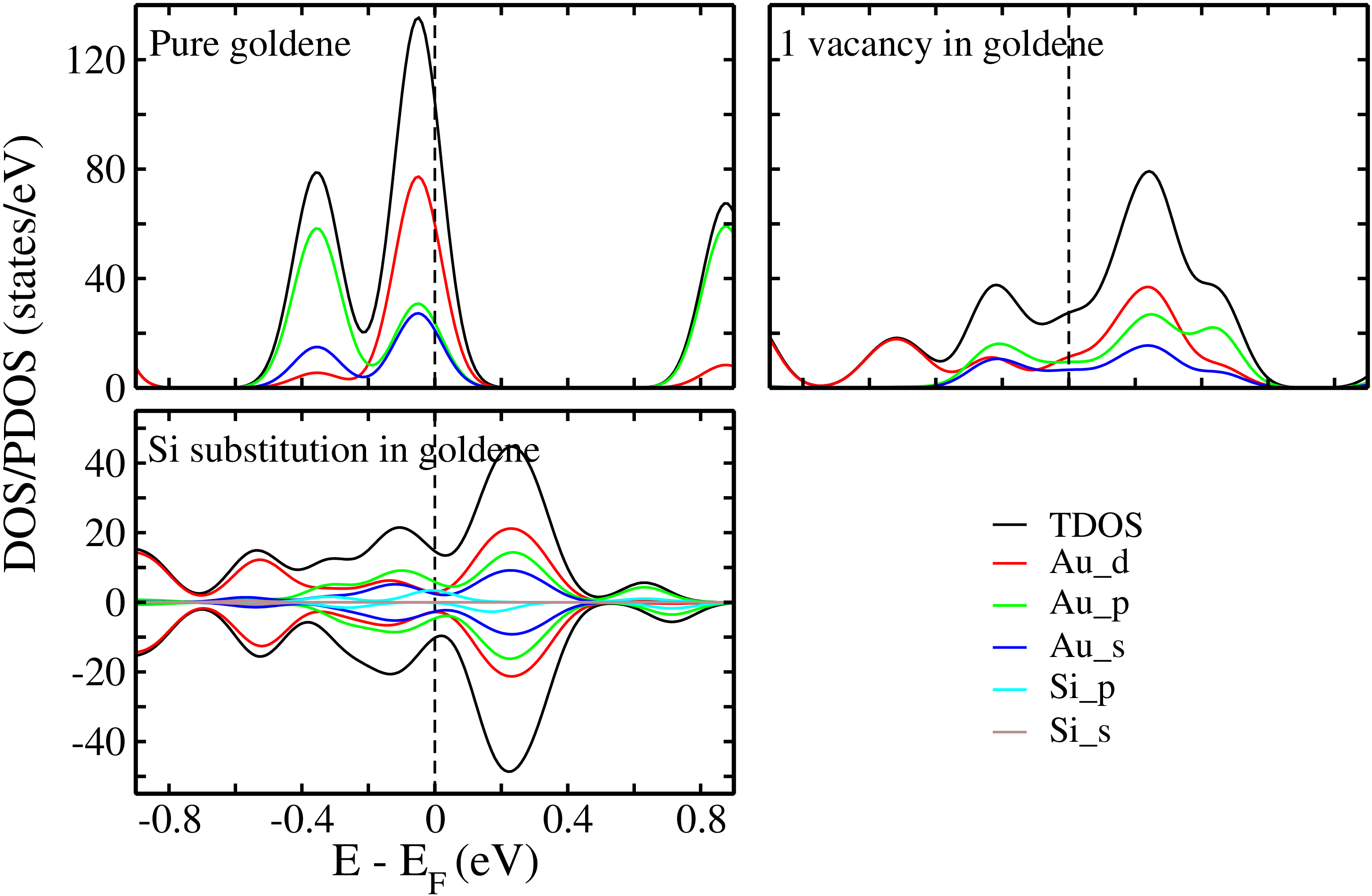}
\caption{Calculated densities of states of pure Au 2D material, one Au-vacancy and Si-doped goldene.}
\label{dossubSi-goldene}
\end{figure}

In Fig.\ref{dossubSi-goldene}, we plot the density of electronic states of pure, defective, and Si-doped goldene. For the doped system, at the Fermi level, the p and s orbitals of the Au atom and the p states of Si show higher involvement, whereas the d, p, and s orbitals of Au contribute to the states above the Fermi level. The p orbitals of silicon contribute to the states below the Fermi level more than those above it. On the other hand, we see that the electron population around the Fermi level decreased from the pure system to the Si-doped structure. We also remark that when one Si atom is doped into the Au 2D material, the system becomes magnetic. The produced magnetic moment is approximately 0.63 $\mu_{{\rm B}}$. The p states of the Si atom mainly contribute to the induced magnetic moment. The local magnetic moment of the Si atom is 0.40 $\mu_{{\rm B}}$. However, Au atoms are slightly magnetized. It is important to note that the dangling bonds associated with Au vacancy, which was created before the adsorption of Si atom, is present in the defective model. Removing one Au atom leaves the surrounding Au atoms unsaturated, generating localized states at the vacancy site; these vacancy-induced dangling bonds before the adsorption, therefore giving raise to localized electronic states at the defect site. These vacancy-induced dangling bonds interact directly with the Si atom and play important role in the magnetism. On the other hand, we note that we have also calculated the energy difference between the magnetic and non-magnetic states to verify the stability of the magnetism.

\begin{figure}[H]
\centering
\includegraphics[width=0.6\textwidth]{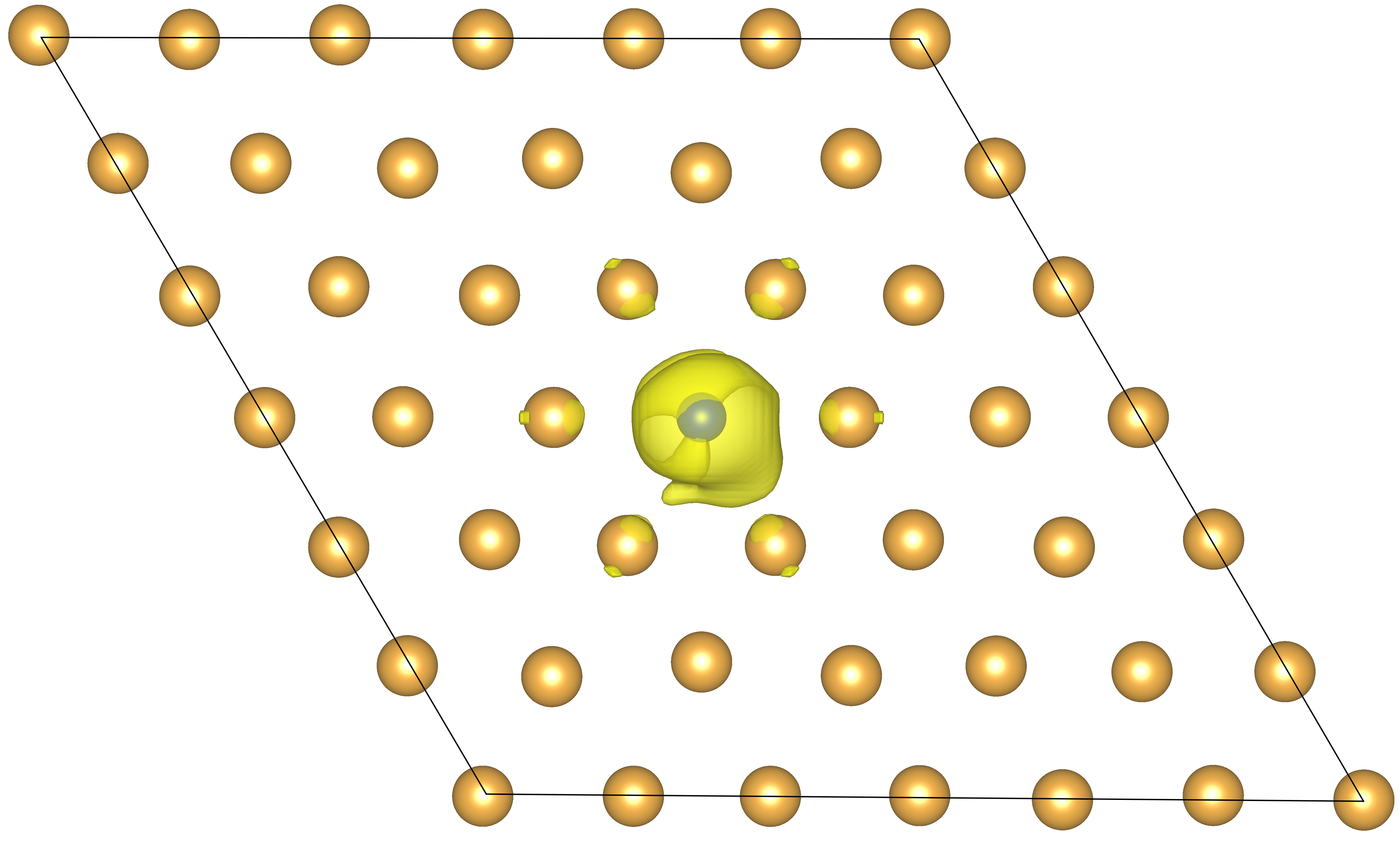}
\caption{Spin density distribution of the Si-doped goldene.}
\label{spindensSi-goldene}
\end{figure}

The spin density distribution of the Si-doped 7$\times$7$\times$1 Au 2D material is shown in Fig. \ref{spindensSi-goldene}.The induced spin polarization is low for Si-doped Au 2D material. Moreover, we can see that magnetism originates mainly from the Si atom, and there is a low magnitude of magnetic moments at the 1$^{{\rm st}}$ NN of the Si atom and in the interstitial area. The interstitial magnetic moments could provide insight that the hybridization between the Si atom and neighboring Au atoms plays a role in the induced magnetism.


\section{Conclusion}
In summary, we have conducted DFTB calculations to study the electronic and transport properties of Au 2D material and the effect of a Si-doped atom on the magnetic properties of the Au monolayer. Phonon and molecular dynamic computations were carried out to demonstrate the thermal and dynamical stability of the present material. Moreover, we calculated the mechanical properties and found that the Au 2D material has a Young's modulus value in the same range as that of graphene and higher than that of MoS$_{{\rm 2}}$, WO$_{{\rm 2}}$, hBN, and SiC. We also combined the DFTB method with the NEGF technique to investigate the electronic transport properties of the Au/Au/Au junction. The junction is electrically conductive. The transmission spectrum of the Au 2D material indicates that it is highly conductive.This result was confirmed by the current-voltage curve, which shows a linear increase. Finally, we studied the effect of Si doping on Au 2D material. We found that the Si atom can induce low-spin states in the goldene monolayer, producing a magnetic moment of 0.63 $\mu_{{\rm B}}$. \\
According to our DFTB studies, goldene 2D material may provide an unusual combination of mechanical robustness, metallic conductivity, and chemical nobility that hardly ever occurs at the same time in known monolayer systems. We are confident that the present results may provide a guide for future investigations, primarily experimentally, to fabricate our proposed 2D-based device, as well as for theoretical studies in search of high quality goldene based junctions.

\begin{acknowledgement}
This article was produced with the financial support of the European Union under the LERCO project (number ${\rm CZ.10.03.01/00/22\_003/0000003}$) {\it via} the Operational Programme Just Transition. The calculations were performed at IT4Innovations National Supercomputing Center through the e-INFRA CZ (grant ID:90254).
\end{acknowledgement}

\begin{suppinfo}

\end{suppinfo}

\bibliography{refs.bib}

\vspace{5cm}
\begin{figure}[H]
\includegraphics[width=0.98\textwidth]{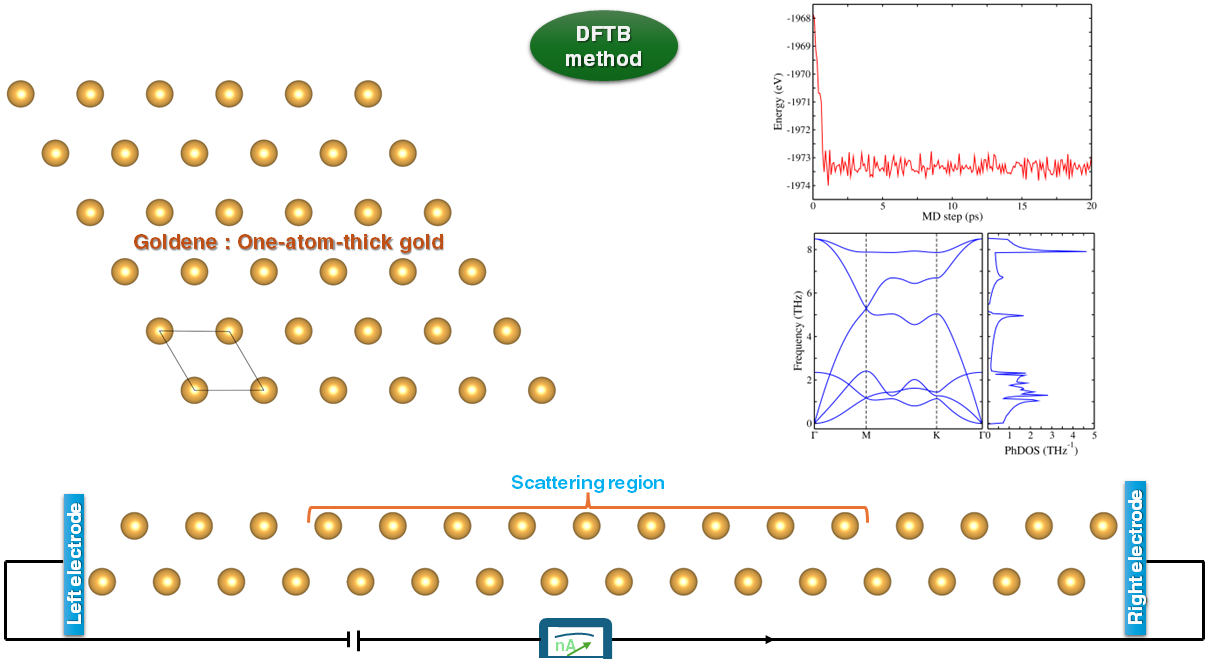}
\caption*{TOC Graphic}
\label{}
\end{figure}

\end{document}